\documentclass[apj]{emulateapj}

\usepackage{bm}
\usepackage{graphicx}
\usepackage{epsf}
\usepackage{graphics}

\def\mpc{\,h^{-1}{\rm Mpc}}

\def\msun{\,h^{-1}{\rm M}_\odot}

\makeatletter

\newcommand{\Rmnum}[1]{\expandafter\@slowromancap\romannumeral #1@}
\makeatother

\shorttitle{halos in the filaments}

\shortauthors{Zhang et al.}

\begin{document}


\title{The spin and orientation of dark matter halos within cosmic filaments}

\author{Youcai Zhang\altaffilmark{1,4}, Xiaohu Yang\altaffilmark{1}, Andreas
  Faltenbacher\altaffilmark{2}, Volker Springel\altaffilmark{2}, Weipeng
  Lin\altaffilmark{1}, Huiyuan Wang\altaffilmark{3} }

\altaffiltext{1}{Key Laboratory for Research in Galaxies and Cosmology,
  Shanghai Astronomical Observatory; the Partner Group of MPA; Nandan Road 80,
  Shanghai 200030, China; E-mail: yczhang@shao.ac.cn}

\altaffiltext{2}{Max-Planck-Institut f\"ur Astrophysik,
  Karl-Schwarzschild-Strasse 1, 85748 Garching, Germany}

\altaffiltext{3}{Key Laboratory for Research in Galaxies and Cosmology, Center
  for Astrophysics, University of Science and Technology of China, 230026,
  P. R. China}

\altaffiltext{4}{Graduate School of the Chinese Academy of Sciences, 19A,
  Yuquan Road, Beijing, China}

\begin{abstract} Clusters, filaments, sheets and voids are the building blocks
  of the cosmic web. Forming dark matter halos respond to these different
  large-scale environments, and this in turn affects the properties of
  galaxies hosted by the halos. It is therefore important to understand the
  systematic correlations of halo properties with the morphology of the cosmic
  web, as this informs both about galaxy formation physics and possible
  systematics of weak lensing studies.  In this study, we present and compare
  two distinct algorithms for finding cosmic filaments and sheets, a task
  which is far less well established than the identification of dark matter
  halos or voids. One method is based on the smoothed dark matter density
  field, the other uses the halo distributions directly.  We apply both
  techniques to one high resolution N-body simulation and reconstruct the
  filamentary/sheet like network of the dark matter density field. We focus on
  investigating the properties of the dark matter halos inside these
  structures, in particular on the directions of their spins and the
  orientation of their shapes with respect to the directions of the filaments
  and sheets. We find that both the spin and the major axes of filament-halos
  with masses $\lesssim 10^{13}\msun$ are preferentially aligned with the
  direction of the filaments. The spins and major axes of halos in sheets tend
  to lie parallel to the sheets. There is an opposite mass dependence of the
  alignment strengths for the spin (negative) and major (positive) axes, i.e.
  with increasing halo mass the major axis tends to be more strongly aligned
  with the direction of the filament whereas the alignment between halo spin
  and filament becomes weaker with increasing halo mass. The alignment
  strengths as a function of distance to the most massive node halo indicate
  that there is a transit large scale environment impact: from the 2-D
  collapse phase of the filament to the 3-D collapse phase of the cluster/node
  halo at small separation. Overall, the two algorithms for filament/sheet
  identification investigated here agree well with each other. The method
  based on halos alone can be easily adapted for use with observational data
  sets.
\end{abstract}

\section{INTRODUCTION}\label{sec_intro}
Inspection of the galaxy distribution in redshift surveys (e.g.~in the Sloan
Digital Sky Survey, York et al. 2000) or of the distribution of dark matter
particles in numerical simulations (e.g.  Millennium Simulations, Springel et
al.  2005) reveals a striking `cosmic web' (Bond et al. 1996), composed of
clusters, filaments, sheets and voids as primary building blocks.  In the
matter or galaxy distributions, one can clearly see large volumes of `empty'
regions (voids) which are surrounded by thin denser sheetlike structures
(sheets).  At even higher density contrast a network of filaments dominates
the matter or galaxy distributions. Finally, the massive clumps at the
intersections of filaments typically correspond to rich galaxy groups and
clusters. The human eye is readily capable of identifying these morphological
features of the cosmic large-scale structure, but for quantitative analysis
objective techniques for structure analysis are needed.

The most commonly employed statistical measure for the distribution of matter
and galaxies are low-order clustering statistics (e.g., the two-point
correlation function and its Fourier transform, the power spectrum; Peebles
1980).  Apart from these standard statistical tools, one may also use other
statistics, for example based on higher-order correlation functions, or
measures for the abundance of halos as a function of mass or of voids of
different sizes, to characterize cosmic structure.

Since halos and voids are approximately spherically symmetric structures and
relatively easy to model, several very successful methods to extract these
structures from simulations and observations have been developed. In N-body
simulations, halos are most commonly found as groups of particles by the
friends-of-friends (FOF) algorithm with a linking length set equal to some
fraction ($b\approx0.2$) of the mean particle separation (e.g. Davis et al.
1985). Halos detected in this way show a mean enclosed density of about $180$
times the average mass density of the universe.  Observationally, galaxy
groups can also be identified using the FOF method, but with two linking
lengths in order to take redshift distortions into account (e.g., Eke et al.
2004).  One may also use models for the density profile and velocity
dispersion of dark matter halos to help extracting galaxy groups (e.g., Yang
et al. 2005). There are numerous successful void finders, many of them have
been compared in detail in a recent study by Colberg et al.~(2009; and
references therein).

Compared to clusters and voids, filaments and sheets are more complicated
geometric structures. Both of them are genuine 3-dimensional objects
associated with a distinct orientation in space. Their density contrast is
often quite close to the mean cosmic density, making their identification
difficult and sometimes ambiguous.  We also note that from a dynamical point
of view there exists a sequence of transformations that can cast one of these
structures into another, which blurs any clear distinction between these
different morphological structures.  Matter tends to flow out of the voids,
accretes onto the sheets, collapses to the filaments, and finally accumulates
onto the large clumps at the intersection of the filaments. These processes
are expected to impact the properties of the halos and the galaxies formed
therein, leading to correlation between halo properties and large-scale
environment.  In a recent study, using high-resolution N-body simulations, Gao
et al.~(2005) found that the halo clustering strength not only depends on the
masses of the halos but also on their formation times (see also Sheth \&
Tormen 2004; Wang et al. 2007; Dalal et al. 2008; Hahn et al. 2009). In
addition to the age, other halo properties such as concentration and spin have
also been found to correlate with the local environment (Avila-Reese et al.
2005; Wechsler et al.  2005; Bett et al.  2007; Gao \& White 2007; Macci\`o et
al.  2007).

Based on semi-analytical galaxy evolution models, Croton et al.~(2007) showed
that besides the properties of the host halo the large scale environment has
to be taken into account to fully characterize the galaxies at halo centers
(see also Reed et al.  2007).  Observations indicate a similar dependence of
galaxy properties on environment.  Using a large galaxy redshift catalog, Yang
et al.  (2006) showed that groups (of similar masses) with red central
galaxies are more strongly clustered than those with blue central galaxies
(see also Wang et al. 2008).  Using the same data set, Wang et al. (2009b)
detected numerous red dwarf galaxies which were isolated, i.e.~which were not
belonging to any neighboring larger association or group of galaxies.  The
origin of those red isolated dwarfs still remains unknown (see also Cooper et
al. 2007).  To shed some more light on the impact of large-scale structure on
halo/galaxy properties we here adopt the following strategy. We first identify
and classify the large-scale environment, then we determine whether a given
halo resides in a filament or sheet, and finally we try to find correlations
between the halo spin and shape orientations and their large-scale
environment.

Up to the present day, a number of different approaches have been suggested to
find filaments (and/or sheets) in simulations as well as in observations.
Among these methods two different classes of techniques can be distinguished:
the first uses discrete distributions of objects (e.g., galaxies) and the
second is based on continuous density fields.  In the following, we briefly
summarize the basic ideas of these two approaches (see also Arag\'on-Calvo et
al.  2007a; and references therein).

A discrete point set allows the use of a minimum spanning tree technique to
link the points (Barrow et al.  1985; Colberg 2007).  However, cosmic
structures identified by this method often show web-like features and it is
difficult to define sizes and directions of the extracted filaments.  Finding
filaments joining neighboring clusters has also been carried out.  Pimbblet
(2005) searched the 2dF Galaxy Redshift Survey catalog for filamentary
structures using the orientations of inter-cluster galaxies. A related
approach based on the inter-cluster dark matter distribution derived from
N-body simulations is discussed in Colberg et al.  (2005). More
mathematically, Stoica et al. (2005; 2007) used a so called {\it Candy model}
to locate the filaments in galaxy surveys. All these approaches have the
advantage that they directly deal with the original point sources and do not
require the creation of continuous density fields.  However, in general these
techniques incorporate a large number of free parameters and the specific
assignment of these parameters is, to a certain degree, arbitrary.

The second type of techniques usually require the creation of a continuous
density field. These approaches include the full characterization of the
topology of the matter distribution in terms of Minkowski functionals (Mecke
et al. 1994; Schmalzing et al. 1999), the genus of the density field (Gott et
al. 1986; Hoyle \& Vogeley 2002) and the shape-finder statistics in three
(Sahni, Sathyaprakash \& Shandarin 1998) and two dimensions (Pandey \&
Bharadwaj 2008). A more rigorous classification of filaments and other
structural elements is obtained with the skeleton analysis of density fields
by Novikov et al. (2006) and Sousbie et al.  (2008), where the maxima and
saddle points in the density field are specified using the Hessian matrix
(i.e. the second partial derivative) of the density field. Recently, Hahn et
al.  (2007a; b) have quantified the cosmic web using the Hessian matrix of the
potential field where according to the number of positive eigenvalues a region
was classified as belonging either to a cluster, filament, sheet or void. The
only free parameter in this analysis is the smoothing length of the density
field.  In a similar spirit, Arag\'on-Calvo et al. (2007a; b) computed the
Hessian matrix based on the density field constructed with a Delaunay
Triangulation Field Estimator (Schaap \& van de Weygaert 2000).  

The methods in this second category are mathematically more rigorous than the
techniques based on discrete point sets. Only a few (or even no) free
parameters are needed, and all halos (galaxies) can be classified since
essentially every volume element and hence the objects within it can be
classified. However, a disadvantage lies in the necessity for constructing a
density (potential) field, and in the difficulties in assigning points, halos,
or galaxies, to the appropriate volume element.

In this study, we compare two types of filament finding algorithms, one using
the overall matter distribution and one just the halos obtained from a
high-resolution cosmological N-body simulation.  Method \Rmnum{1} employs the
Hessian matrix of the mass density field, where halos are differentiated into
four types using a combination of criteria from Hahn et al. (2007a) and
Arag\'on-Calvo et al. (2007a).  Method \Rmnum{2} uses the halo distribution
directly. Our development of this approach has been inspired by the Candy
model approach suggested by Stoica et al.~(2005) and the connecting cluster
technique described in Colberg et al.~(2005).  Based on either method we
extract filaments and sheets from our high resolution N-body simulation. We
then correlate the characteristics of the cosmic structures, like sizes and
orientations, with the spins and orientations of dark matter halos.

This paper is organized as follows.  In Section \ref{sec_data}, we give a
brief description of our N-body simulation and the halo catalog we used.  We
present a detailed description of the filament-finding methods based on either
the dark matter density field or the distribution of dark matter halos, in
Section \ref{sec_method}.  In Section~\ref{sec_results}, we analyze various
alignment signal measurements for halos with respect to the filaments (and
sheets).  Finally, we summarize our results in Section~\ref{sec_summ}.
\section {N-body simulations and the halo catalog}\label{sec_data}
\subsection{Simulation parameters and halo definition}
\begin{figure*}
  \plotone{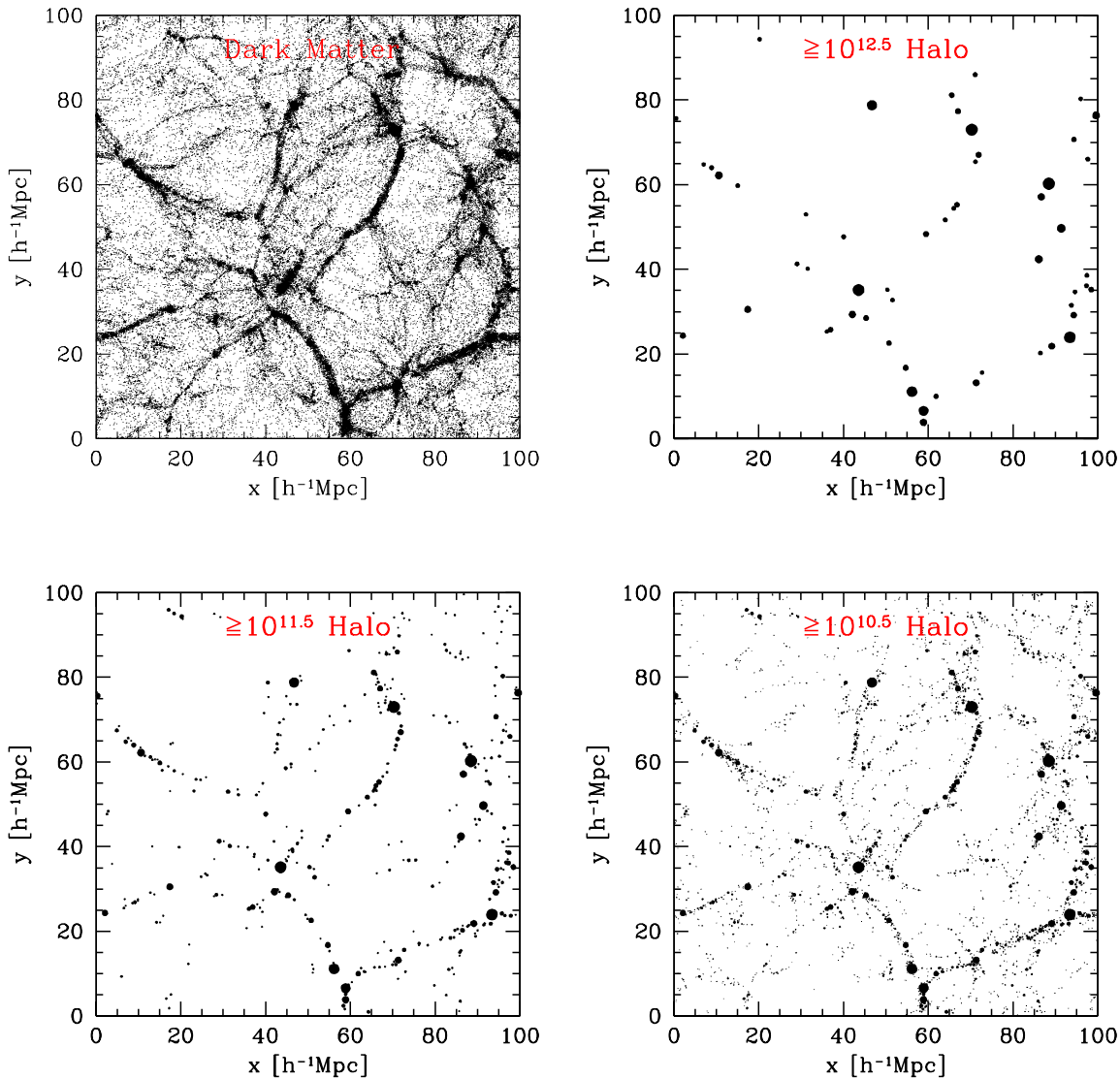}
  \caption{The dark matter (upper-left panel) and halo distributions in a
    slice of thickness $4\mpc$ through the full box ($100\mpc$ on a side). For
    the halo distributions, the halo mass range (lower limit) is indicated in
    each panel, and the sizes of the dots are proportional to the virial radii
    of the halos. }
\label{fig:simulation}
\end{figure*}
In this study, we make use of a N-body simulation carried out at the Shanghai
Supercomputer Center using the massively parallel GADGET2 code (Springel et
al. 2001; 2005).  The simulation evolved $1024^3$ dark matter particles in a
periodic box of $100\mpc$ on a side from redshift $z=120$ to the present
epoch. The particle mass and softening length are $6.927\times10^7\msun$ and
$2.25{\,h^{-1}{\rm kpc}}$, respectively. The cosmological parameters used in
the simulation are $\Omega_{\rm m}= \Omega_{\rm dm}+\Omega_{\rm b}=0.25$,
$\Omega_{\rm b}=0.045$, $h=0.73$, $\Omega_\Lambda=0.75$, $n=1$, and
$\sigma_8=0.9$. In the upper-left panel of Fig.~\ref{fig:simulation}, we show
the distribution of dark matter particles in a slice of thickness $4\mpc$ at
redshift $z=0$. For clarity, only 0.15\% of the dark matter particles are
plotted.  This representation clearly shows the well-known features of the
non-linear cosmic density field, in particular the prominent filamentary
structures that coined the term `cosmic web' are nicely seen.

Dark matter haloes were identified from the simulation at redshift $z=0$ using
the standard friends-of-friends (FOF) algorithm (Davis et al. 1985) with a
linking length of 0.2 times the mean interparticle separation.  Haloes
obtained with this linking length have a mean over-density of $\sim 180$
(Porciani, Dekel \& Hoffman 2002).  As an illustration, we show in the
upper-right, lower-left and lower-right panels of Fig.~\ref{fig:simulation}
the halo distributions for a sequence of decreasing lower mass limits: $\ge
10^{12.5}$, $\ge 10^{11.5}$ and $\ge 10^{10.5}\msun$, as indicated in the
panels. The sizes of the dots are scaled to be proportional to the virial
radius of the dark matter halos. These plots indicate that the distribution of
halos with masses $\ge 10^{12.5}\msun$ can only resolve the high density
regions at the nodes of the cosmic web, while halos with masses $\ge
10^{11.5}\msun$ trace the filamentary structures quite well. Including smaller
halos down to $\ge 10^{10.5}\msun$ can reveal subtle features even in
void-like regions. As discussed in Yang et al.  (2009; Fig. 1), the SDSS
observations can completely resolve the dark matter halos with a mass limit of
$\ge 10^{12.5}\msun$ at redshift $z\sim 0.12$, and of $\ge 10^{11.5}\msun$ at
redshift $z\sim 0.05$. Thus the halo-based probe for filamentary structures
investigated in this paper can be reliably applied to the SDSS observations at
low redshifts.

The main purpose of this paper is to probe the orientations and spins of dark
matter halos with respect to the filaments and sheets within the cosmic web.
However, a reliable measurement of these halo properties, in particular of the
spin, requires high mass resolution. Therefore we only retain halos with at
least $500$ particles for further analysis, resulting in a catalog with
$73068$ halos.

\subsection{Measuring the spin and orientation of the halos}

The angular momentum $\boldsymbol J$ of a FOF halo containing $N$ particles is
\begin{equation}
\boldsymbol{J}=m\sum_{i=1}^N \boldsymbol{r_i} \times \boldsymbol{v_i},
\end{equation}
where $m$ is the particle's mass, $\boldsymbol{r_i}$ is the position vector of
the $i$-th particle relative to the center of  mass,  and $\boldsymbol{v_i}$ is
its velocity relative to the bulk velocity of the halo.

In order to determine the shape of a FOF halo, we use the moment of inertia
tensor\footnote{Actually, Eq.~(\ref{eq:inertia}) represents the second moment of
  the mass tensor. The correct definition of the moment-of-inertia tensor
  deviates from this expression (see e.g., Eq.~1 in Hahn et al.~2007a).
  However, in the context of cosmological alignment studies it has become
  a convention to call the second moment of mass tensor inertia tensor.}
$\boldsymbol I$ with
\begin{equation} 
\label{eq:inertia}
I_{\alpha\beta}=m\sum_{i=1}^N x_{i,\alpha}x_{i,\beta},
\end{equation}  
where $x_{i,\alpha}$ denotes the component $\alpha$ of the position vector of
particle $i$ with respect to the center of mass. The axis ratios $a$, $b$ and
$c$ $(a\ge b \ge c)$ are proportional to the square roots of the corresponding
eigenvalues, $\lambda_1$, $\lambda_2$ and $\lambda_3$, and the orientation of
the halo is determined by the corresponding eigenvectors.
\section{Finding filaments in the N-body simulation}\label{sec_method}
In this  section, we investigate two  distinct methods to find  the filaments in
the  simulation: one  using the  mass density  field and  the other  using the
distribution of dark matter halos.
\subsection{Density field Hessian matrix method}

The first method we examine, hereafter referred to as Method \Rmnum{1}, is
based on the local Hessian matrix $H$ of the smoothed mass density field,
defined as
\begin{equation}\label{HessianMatrix}
  H_{\alpha \beta}=\frac {\partial^2\rho_s(\boldsymbol{x})} 
  {\partial x_\alpha \partial x_\beta},
\end{equation}
where $\rho_s(\boldsymbol{x})$ is the smoothed density field.  $\alpha$ and
$\beta$ denote the Hessian matrix indices with values of 1, 2 or 3. Thus, at
the location $\boldsymbol{x}$ of each halo we can quantify the local ``shape''
of the density field by calculating the eigenvalues of the Hessian matrix
(e.g., Arag\'on-Calvo et al. 2007a). The number of positive eigenvalues of
$H_{\alpha \beta}$ can be used to classify the possible environments in which
a halo can reside into four regions, according to:
\begin{itemize}
\item $cluster$: a region with no positive eigenvalue;
\item $filament$: a region with one positive and two negative eigenvalues;
\item $sheet$: a region with two positive and one negative eigenvalues; 
\item $void$: a region with three positive eigenvalues.
\end{itemize} 
The direction of a filament can be identified with the eigenvector
corresponding to the single positive eigenvalue of the Hessian matrix in a
filament region. We note that this method has only one free parameter, namely
the smoothing scale $R_s$ of the Gaussian filter employed in constructing the
density field. Different from the Multi-scale Morphology Filter (MMF) method
discussed in Arag\'on-Calvo et al. (2007a), we follow Hahn et al.~(2007a) and
adopt a fixed smoothing length of $R_s=2.1\mpc$.

Hahn et al. (2007a) pointed out that the relative volume fractions of the four
categories depend on the choice of $R_s$. For $R_s=2.1\mpc$     (corresponding
to  halo masses of $\sim  10^{13}\msun$) they find a  good agreement between  their
orbit-stability    criterion  and a  visual classification  of the large-scale 
structure.\footnote{Note, however, that  Hahn  et  al.  compute the  Hessian
Matrix  based  on  the gravitational potential field, $\phi$, which is derived
from the matter density distribution via Poisson's equation,
\begin{displaymath}
\bigtriangledown^2\phi = 4\pi G \rho_s(\boldsymbol{x}) \,,
\end{displaymath} 
where  the Poisson's  equation is  solved using  a fast  Fourier  transform on
$1024^3$ grid cells. As a consequence of that,  their eigenvalues have opposite
signs compared to those based on the corresponding density field that we use here. 
}

\begin{figure*}
  \plotone{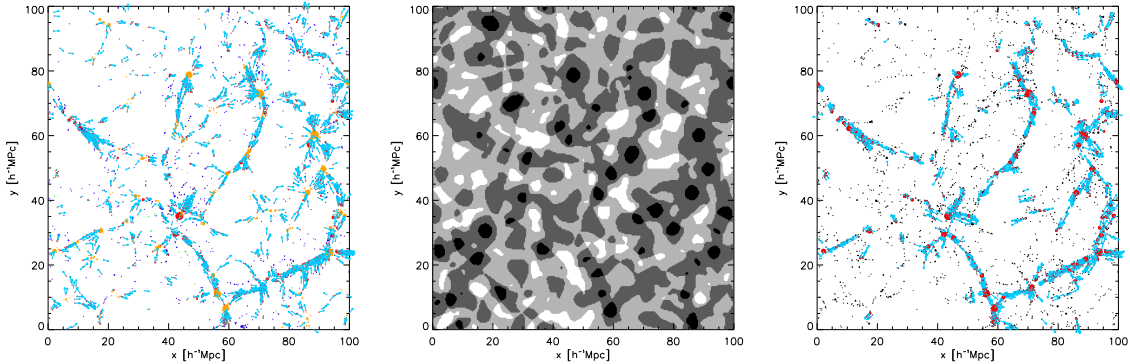}
  \caption{Halo distributions and environmental classifications. The left
    panel is similar to the lower-right panel of Fig.  \ref{fig:simulation},
    but for halos with at least 500 particles (i.e.  $\ge 10^{10.54}
    \msun$). The halos in four different environments are classified by
    different colors: clusters (orange), sheets (blue), filaments (red), voids
    (green).  The cyan arrow indicates the direction of the filament at the
    center of each halo.  The direction of the filament is given by the
    eigenvector corresponding to the single positive eigenvalue of the density
    Hessian matrix. The middle panel shows the environmental classifications
    of $256^2$ grid points in the middle plane of the slice. Black corresponds
    to clusters, dark gray to filaments, clear gray to sheets, and white to
    voids. The right panel is similar to the left panel, but shown for
    filamentary structures identified using the segment extraction method
    (Method II).  The halos in the filaments (both node and member halos) are
    plotted using red color. In addition, each member halo is also marked with
    an arrow indicating the direction of the filament it resides in. }
  \label{fig:fila1}
\end{figure*}

In order to compute the Hessian matrix $H_{\alpha \beta}$, we first construct
a continuous density field from the discrete distribution of particles in the
N-body simulation, using the cloud-in-cell (CIC) technique with a $1024^3$
grid.  Then the smoothed density field $\rho_s$ is produced by smoothing the
CIC generated density field $\rho_{\rm cic}$ with a spherically symmetric
Gaussian filter $G_{R_s}$,
\begin{equation}\label{SmoothDensity}
\rho_s(\boldsymbol{x})=\int {\rm d} \boldsymbol{y}\, \rho_{\rm cic}(\boldsymbol{y})
G_{R_s}(\boldsymbol{y},\boldsymbol{x}),
\end{equation}
where $\boldsymbol{x}$ corresponds to the location of a given halo     and the
Gaussian
filter with smoothing scale, $R_s$, is given by
\begin{equation}\label{GassianFilter}
G_{R_s}=\frac {1}{(2\pi R_s^2)^{3/2}} \exp
\left( -\frac{|\boldsymbol{y}-\boldsymbol{x}|^2} {2R_s^2}\right).
\end{equation}
From     equations     (\ref{HessianMatrix}),    (\ref{SmoothDensity})     and
(\ref{GassianFilter}), we find
\begin{eqnarray}
H_{\alpha\beta}&=&\frac{1}{R_s^4}\int {\rm d} \boldsymbol{y} \,
\left[ (x_\alpha - y_\alpha)(x_\beta - y_\beta) - \delta_{\alpha \beta}R_s^2
\right] \nonumber\\
&&\rho_{\rm cic}(\boldsymbol{y}) G_{R_s}(\boldsymbol{y},\boldsymbol{x}),
\end{eqnarray}
where  $\delta_{\alpha  \beta}$  is  the Kronecker  delta  (Arag\'on-Calvo  et
al. 2007a). Finally, the eigenspace  structure of the symmetric Hessian matrix
can be computed at the center of mass  of each halo. 

According to the number of positive eigenvalues at the locations of the dark
matter halos we classify them into four categories, as outlined above. The
numbers of halos in cluster, sheet, filament and void regions are,
respectively, $13803$, $13230$, $45755$ and $280$, corresponding to $18.89$,
$18.11$, $62.62$ and $0.38$ per cent of the total number of halos. In the left
panel of Fig.~\ref{fig:fila1}, we show the distribution of halos in a slice of
thickness $4\mpc$, indicating the classification according to their
environments with different colors.  In addition, the directions of the
filaments are marked with arrows for those halos associated with them. The
right panel of Fig.~\ref{fig:fila1} shows the environmental classification of
$256^2$ grid point put down regularly in the mid-plane of the slice. According
to visual inspection, the general appearance of the filamentary structures
identified with this method is remarkably good.

\subsection{Segment extraction method}
\begin{figure}
  \plotone{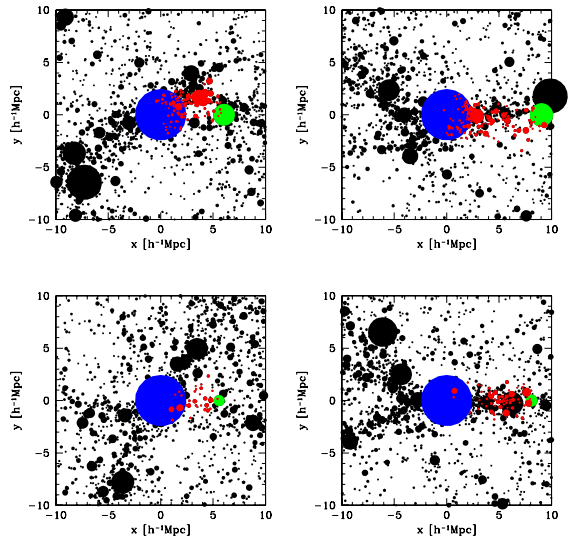}
  \caption{Distribution of dark  matter halos around the most  massive halo in
    our simulation, where only halos with  at least 500 particles and within a
    $(20\mpc)^3$ box centered  at the most massive halo  are plotted.  In each
    panel,  the  colored halos  are  within  one  filamentary segment:  (blue:
    starting  node halo;  green: ending  node halo;  red: member  halos). Four
    segments in a total of six associated with the most massive halo are shown
    in four panels. For better  visual quality, the distributions of halos are
    rotated so that the segment is always displayed along the $x$-axis. }
\label{fig:test}
\end{figure}
\begin{figure}
  \plotone{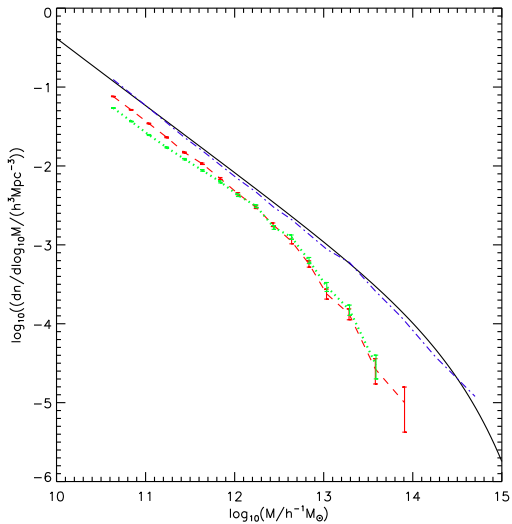}
  \caption{Mass function of the FOF halos residing in filaments in Method
    \Rmnum{1} (red dashed) and Method \Rmnum{2} (green dotted). Poisson error
    bars are added to assess the difference between the two methods. The
    filament mass function is scaled to the whole volume of our simulation.
    For comparison purposes, we also show the total mass function from our
    simulation (blue dash-dotted) and SMT analytic model prediction (black
    solid).}
\label{mass_function}
\end{figure}

As shown in the previous section, filamentary structures can be well
identified using the Hessian matrix of the density field. However, this
approach relies on an accurate knowledge of the matter density field, which is
in many cases highly non-trivial to obtain.  In particular, for observational
data one may wish to use the distribution of galaxies, groups or halos
directly to identify the filaments of the cosmic web.  For this purpose, we
now present a halo-based method for filament finding which is based on a
slightly modified version of the `Candy model' proposed by Stoica et
al.~(2005).  The Candy model reconstructs filaments by connecting individual
segments that are found in a basic point distribution (galaxies, halos, etc.).
In this study, we only aim to compare halo shape orientations with the
orientations of the segments they are residing in, thus we do not discuss in
detail the problem of composing individual segments into long connected
filaments.  Hereafter, the segment extraction method derived from the Candy
model will be referred to as Method \Rmnum{2}. In brief, there are two
fundamental differences between Method \Rmnum{1} and \Rmnum{2}. One is the use
of a biased tracer of the density field in Method \Rmnum{2}. The other is that
halos in Method \Rmnum{2} are automatically grouped into given filaments. That
is, we know what halo belongs to what filament.

The requirements which a group of points has to fulfill to be considered a
candidate segment have to be adjusted to the problem in question.  Here we
focus on dark matter halos as building blocks for the filamentary structure of
the cosmic web.  Therefore, a candidate segment is assumed to be a cylinder,
with a length in the range of $[L_{\rm min}, L_{\rm max}]$ and a radius in the
range $[R_{\rm min}, R_{\rm max}]$. The mean mass density within the segment
should be at least $N_\rho$ times that of the average mass density of all
halos, $\bar{\rho}_h=M_{\rm h,tot}/V$, where $M_{\rm h,tot}$ is the total mass
of the halos with more than $500$ particles and $V$ is the volume of the
simulation.  Finally, a segment should have at least $N_{\rm min}$ member
halos.  We set these free parameters as follows:
\begin{eqnarray}
&&  L_{\rm min} = 3\mpc \nonumber~~~~~~~~~~~~~~~~~~~~~~~~~~~~~~~~~~~~~~~~~\\
&&  L_{\rm max} = 10\mpc \nonumber\\
&&  R_{\rm min} = 1\mpc \nonumber\\
&&  R_{\rm max} = 3\mpc \nonumber\\
&&  N_\rho = 5 \nonumber\\
&&  N_{\rm min} = 5
\end{eqnarray}

Note that the values chosen for these parameters are somewhat arbitrary, but
the results are robust to substantial changes in these parameters. In
particular, we tested a number of different reasonable sets of parameter
values, all of them led to general agreement between the results.
Specifically, we found that changing any of the above listed parameters
(except for $N_\rho$) by 50\% will only result in a less than 5\% change in
the number of halos identified as belonging to filaments. If $N_\rho$ is
varied to $N_\rho=2.5$ or $N_\rho=10.0$, about 59\% or 31\% of the halos,
respectively, are classified as belonging to filaments. Even for these two
cases, the final results do not change qualitatively, however.

In the following we describe the successive steps of the segment
extraction method.

{\bf Step 1.} The dark matter halos are ranked according to their masses.

{\bf Step 2.}   Starting from the most massive halo to  ever smaller ones, we search
around each halo $i$  (node) all other halos $j$  with distance in the
range [$L_{\rm min}, L_{\rm max}$]. These halo pairs form candidates for our
filament segments.

{\bf Step  3.} We calculate for  each candidate segment  the filament strength
(average mass density), ${\cal S}_{i,j}$, which is defined as
\begin{equation}\label{strength}
{\cal S}_{i,j}= \sum_{k=1}^N M_{k}/(\pi R_s^2L_{i,j}) \,.
\end{equation}
Here, $L_{i,j}$ is  the length of the candidate  segment (distance between the
two halos $i$ and  $j$), $N$ is the total number of  halos within the cylinder
around this  segment with  radius $R_s$, and $M_k$ is the  mass of halo  $k$. The
candidate  segment radius,  in  the range  $R_s  \in [\max(R_{\rm  min},r_{\rm
vir,i}), R_{\rm  max}]$, is determined such that ${\cal  S}_{i,j}$ reaches
its maximum.

{\bf Step 4.}  For all the  candidate segments, we rank their strengths ${\cal
  S}_{i,j}$. The  one with the largest  strength, with at  least $N_{\rm min}$
member halos  and ${\cal S}_{i,j}\ge  N_\rho \bar{\rho}_h$, is defined  as the
first candidate segment associated with node halo~$i$.

{\bf Step  5.} We rank all the  member halos within that  segment according to
their masses, except the node halo $i$.  The most massive one $l$ with at least
$N_{\rm min}-2$  halos residing between  halos $i$ and  $l$ is defined  as the
second, terminal  node of the segment.  We remove those member  halos that are
between  the  two  node  halos  (belonging  to segment  $i,l$)  from  the  halo
list. Thus member  halos can only belong to one segment,  while node halos can
belong to more  than one segment.  The direction of  the segment (filament) is
defined to be the connecting line between the two node halos.

{\bf Step 6.} Once the segment $(i,l)$ is determined, we return to step 1 and
search for other segments associated with node halo $i$ until no further
segments can be found for this node halo. As an illustration, we show in
Fig.~\ref{fig:test} the halo distribution in a $(20\mpc)^3$ cube centered on
the most massive halo $i=1$, together with the node and member halos in
segment $(i,l)$ which are marked with colored dots. The first four segments
associated with node halo $i$ (6 in total) are illustrated using colored dots
in the four panels of Fig.~\ref{fig:test}.

{\bf Step  7.} We  turn to  the next node  halo to  search for its associated
segments.  We  iteratively search the  halo catalog until no  further segments
can be found.

Applying the segment extraction method to our halo catalog we find that 45\%
of all halos are classified as members of segments, and 2\% as node halos.
These two populations are substantially smaller than those obtained with
Method \Rmnum{1}, where we found 63\% and 19\% to be classified as filament
and cluster halos, respectively.  Note, however, in a recent paper
Forero-Romero et al.~(2008) introduced an additional threshold $\lambda_{\rm
  th}$ for the classification of environment instead of just using the number
of positive (negative) eigenvalues.  Increasing this barrier from
$\lambda_{\rm th}=0$ to some finite value greatly reduces the volume (mass)
occupied by the identified filaments, and in particular, reduces the number of
ambiguous detections of feeble or spurious filaments. The discrepancy may
hence just be a result of the higher detection threshold of Method \Rmnum{2}
compared with Method \Rmnum{1}. As a more quantitative comparison, in
Fig.~\ref{mass_function}, we show the mass functions of filament halos derived
with Method \Rmnum{1} (red dashed) and Method \Rmnum{2} (green dotted). The
mass functions obtained from the two methods are very similar, especially at
the high-mass end. While at the low mass end, because of the higher detection
threshold of Method \Rmnum{2}, less halos are specified as belonging to the
filaments. For reference, we also show the total mass function from our
simulation (blue dash-dotted), which is in good argeement with the analytic
model prediction by Sheth et al.  (2001; SMT) (black solid).

The right panel of Fig.~\ref{fig:fila1} shows the distribution of halos within
the filaments obtained with Method \Rmnum{2} in a slice of thickness $4\mpc$.
Compared to the filaments detected by the Hessian matrix approach in the left
panel, only strong filaments are extracted by this method, while less
prominent structures are not identified. Yet, the advantage of this method is
that it does not require knowledge of the density field.  This is a
significant advantage especially in the context of observations, where density
reconstruction is often difficult.

%
%
%

%
\section{Orientations of dark matter halos relative to filaments and sheets} 
\label{sec_results}
To quantify the impact of filaments and sheets onto halos, we investigate two
types of alignment signals: one based on a halo's spin and the other on its
orientation.  These two vectors correlate with the orientation of the filament
or the normal vector of the sheet in which the halo resides. Finally, we also
compare the results of the two filament-finding methods considered here. For
simplicity, we will often use the terms `shape' and `spin', as well as
`filament' and `sheet', as an indication of direction in the obvious sense.

\begin{figure}
  \plotone{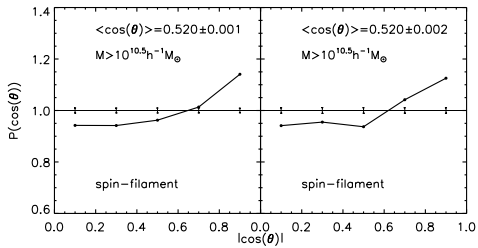}
  \caption{The probability distribution of the cosine of the angle between the
    halo angular momentum vector and the direction of the filament in Method
    \Rmnum{1} (left panel) and Method \Rmnum{2} (right panel). The error bars
    are computed from 500 random samples in which we randomize the
    orientations of the angular momenta of the halos.  In case the angular
    momenta are randomly oriented, we would expect to find $P(\cos \theta
    )=1$.  Thus the error bars are plotted on top of this $P(\cos \theta )=1$
    line.  We also calculate the average value of $\cos(\theta)$, and its
    error, which are given in each panel.}
\label{spin_filament}
\end{figure}
\begin{figure*}
  \plotone{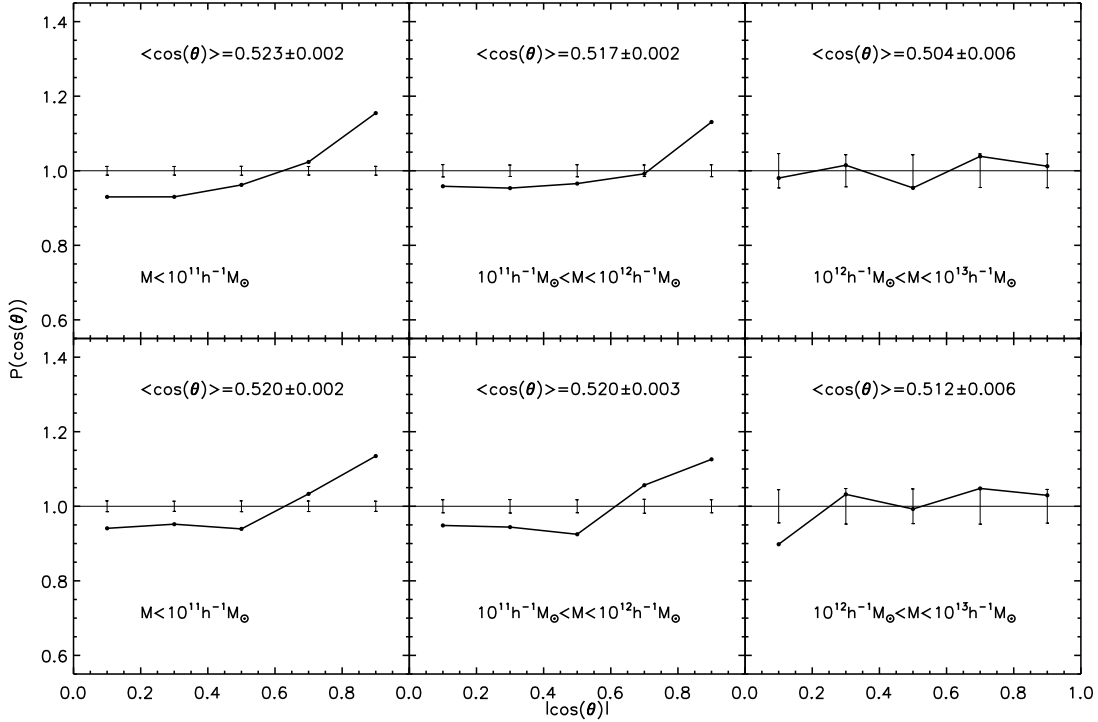}
  \caption{Same as Fig. \ref{spin_filament} but for different mass ranges of
    halos in the filaments in Method \Rmnum{1} (upper panels) and Method
    \Rmnum{2} (lower panels). The average value of $\cos(\theta)$ and its
    error in each mass bin is indicated in each panel.}
\label{spin_massbin}
\end{figure*}
\begin{figure}
  \plotone{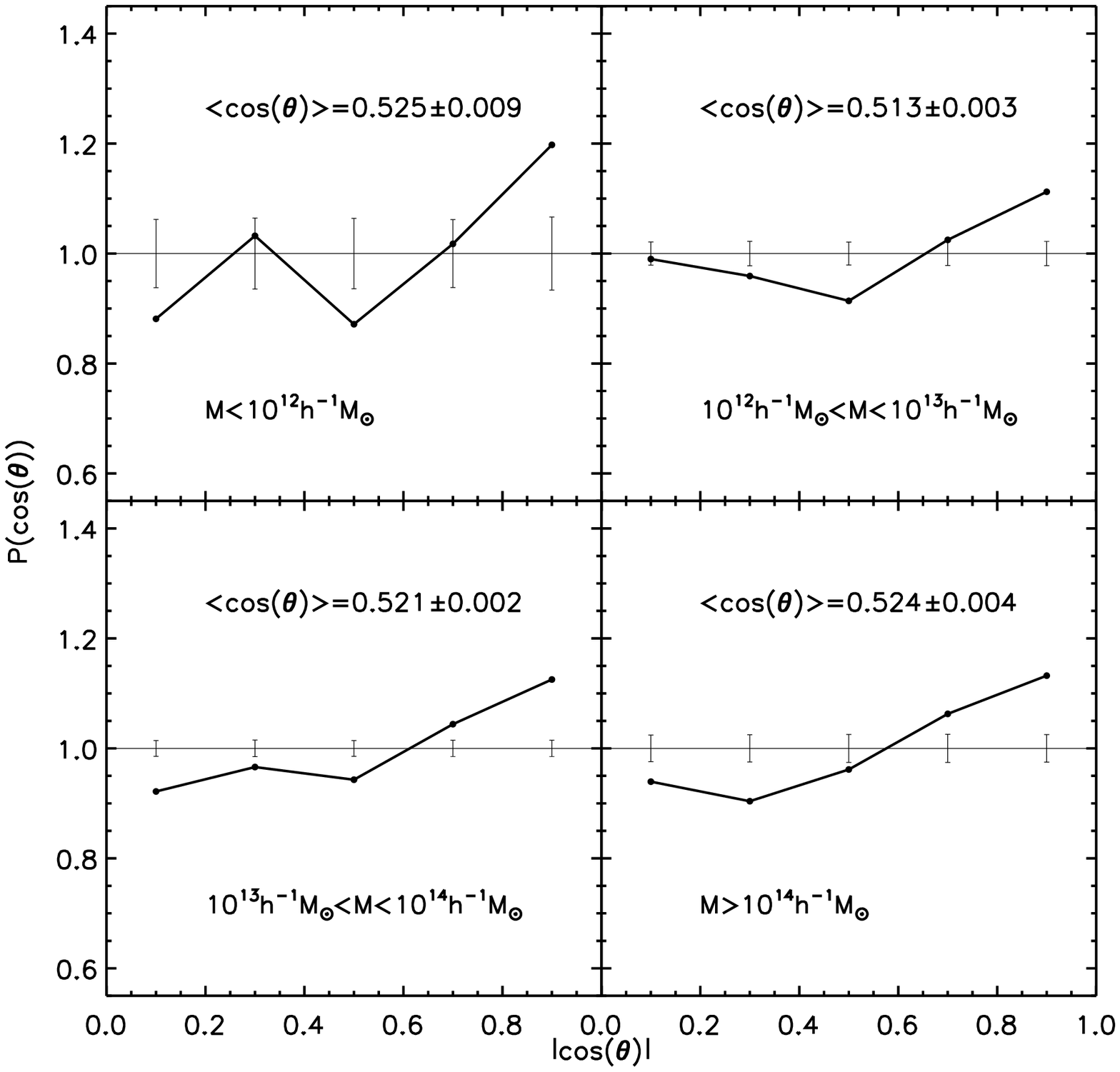}
\caption{Same as Fig.~\ref{spin_filament},  but for halos in different segments
  in Method  \Rmnum{2}, separated  according to the  mass of the  most massive
  halo in each segment. }
\label{spin_masstop}
\end{figure}
\begin{figure}
  \plotone{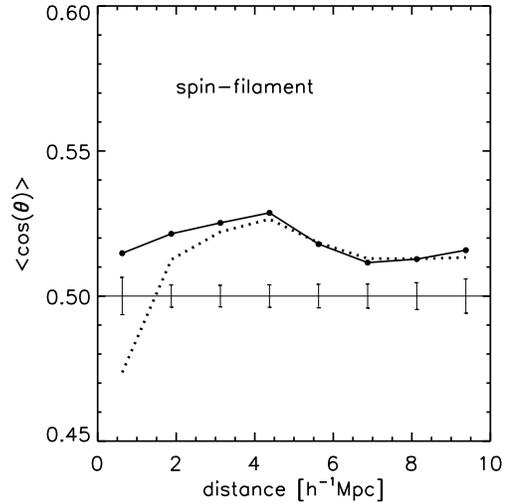}
  \caption{Mean cosine of the angle between spin and filament (solid line),
    between spin and direction of the most massive node halo (dotted line), as
    a function of the separations to the massive node halos in Method
    \Rmnum{2}.}
\label{spin_distance}
\end{figure}
\subsection{Statistical methodology}
In order to quantify the spin and shape orientations of halos relative to
filaments, we compute the probability distribution function $P(\cos \theta)$
(alignment signal), where $\theta$ is the angle between the orientations of
halo and filament or sheet.
\begin{equation}
P(\cos \theta) = N(\cos \theta) / \langle N_{\rm R}(\cos\theta) \rangle
\end{equation}
Here, $\cos(\theta$) is restricted to the range $[0, 1]$, and $\cos(\theta)=1$
implies that the halo orientations are parallel to the filaments while
$\cos(\theta)=0$ indicates perpendicular orientations.  For the null
hypothesis of random orientations of halos relative to the environment one
expects $P(\cos\theta) =1$.

To assess the Poisson sampling errors in our alignment signals, we generate
500 random samples in which the orientations of filaments are kept fixed, but
the halo spin or major axes orientations are randomized. For each of these
samples we compute the $P_{\rm R}(\cos \theta)$, which we use to compute the
significance of any detected alignment signal \footnote{Since the significance
  of the alignment singals is quantified with respect to the null hypothesis,
  that is at which level it deviates from no alignments, throughout this paper,
  we plot the error bars on top of $P_{\rm R}(\cos \theta)=1$ lines. }. In
addition, we also calculate the mean cosine $\langle \cos(\theta) \rangle$ of
the alignment angle.  In the absence of any alignment, $\langle \cos(\theta)
\rangle=0.5$.  The significance of any alignment can be assessed in terms of
$\langle \cos(\theta) \rangle$ and $\sigma_{\cos(\theta)}$, which is the
standard deviation of $\langle \cos(\theta) \rangle_{\rm R}$ for the 500
random samples.

\subsection{Spin-filament alignment}

We first examine possible alignment signals between the spin of halos and the
filaments they reside in.  Method \Rmnum{1}, which is based on the Hessian
matrix, allows to distinguish cluster, sheet, filament and void halos. For
filament halos, the eigenvector of the positive eigenvalue of the Hessian
matrix at the location of the halo indicates the direction of the filament.
Method \Rmnum{2} determines the direction of the filament for each member halo
in a segment as the connecting line between the two terminal node halos of
that segment.

Fig.~\ref{spin_filament} shows the probability distribution $P(\cos \theta)$
of the cosine of the angle between the halo spins and the filaments. The left
and right panels display the results for Methods \Rmnum{1} and \Rmnum{2},
respectively.  We find that the halo spins tend to lie parallel to the
filaments. The results are robust against the filament detection method. It
has been argued that the angular momentum of halos originates from the tidal
field exerted by the surrounding dark matter distribution (Peebles 1969;
Doroshkevich 1970; White 1984). The spin directions are expected to be
preferentially aligned with the planes of the sheets and the directions of the
filaments (e.g., Lee 2004), although hydrodynamical simulations have also
suggested that the spin axis may aline with the intermediate axis at
turnaround (Navarro et al. 2004).

In order to characterize the strength of the alignment between angular momenta
and filaments we calculate their average cosine, $\langle \cos(\theta)
\rangle$. These average values together with their errors are displayed in the
panels of Fig.~\ref{spin_filament}. Although the alignment signals for the two
methods are slightly different, the overall strength of the alignment detected
with Methods \Rmnum{1} and \Rmnum{2} agrees well.

Another interesting question is whether the alignment signal and strength
depend on the mass of the filament halos. Fig.~\ref{spin_massbin} presents the
probability distribution $P(\cos \theta)$ for filament halos in different mass
ranges. Results are shown for Methods \Rmnum{1} and \Rmnum{2} in the upper and
lower panels, respectively.  The overall alignment signals obtained from the
two methods are very similar.  There is a weak mass dependence, in the sense
that the alignment is somewhat weaker for massive halos.  According to the
values of $\langle \cos(\theta) \rangle$, filament halos with masses
$M>10^{12}\msun$ are consistent with being randomly oriented at a 2-$\sigma$
confidence level.

In a recent study, Hahn et al. (2007b) reported that the spin of halos with
mass greater than the characteristic halo mass tends to lie perpendicular to
the host filaments. This trend has been confirmed by Arag{\'o}n-Calvo et al.
(2007b). In a study comparing simulations and observations Paz et al. (2008)
found an indication of this behaviour based on SDSS data. The characteristic
mass for gravitational collapse at redshift $z=0$ is $9.57\times10^{12}\msun$,
calculated for the cosmological parameters used in our simulation.  Due to the
small box size of our simulation, we find less than 100 filament halos with
mass $\ge 10^{13}\msun$, independent of the filament finding method applied.
Most of these massive halos are classified as clusters or node halos. Owing to
the sparse number of filament halos, the statistics is too poor to obtain a
robust measurement of the alignment signal, especially with respect to a
possible transition from alignment to anti-alignment. However, for the well
constrained halo mass ranges $\le 10^{13}\msun$, our alignment signals and
strengths are in very nice agreement with those found by Hahn et al.~(2007b),
and slightly larger than those predicted by Arag{\'o}n-Calvo et al.~(2007b).

In Method \Rmnum{2}, the filaments are defined via segments extracted from the
distribution of dark matter halos.  Associated with each segment, there are
two node halos, one of which is the most massive one among all the associated
halos.  Thus we can probe the alignment signals separately for halos in
filaments with different most massive node halos.  In Fig.~\ref{spin_masstop},
we show the results for four mass bins. There is a hint for a very weak
positive mass dependence on the most massive halo in the segment, but this is
statistically not significant.  In addition to the mass dependence itself, we
can further investigate the alignment signals at different separations to the
most massive halo. The solid line in Fig.~\ref{spin_distance} displays the
alignment signals, the mean cosine of the angle between the spin and filament,
for halos at different distances. We find that the strength of the alignment
is slightly suppressed if the halos are either very close to or far away from
the most massive node halos.  To better understand this, we also measure the
alignment signals for the mean cosine of the angle between spin and direction
of the most massive node halo. The results are shown in
Fig.~\ref{spin_distance} as the dotted line. At large separation, since the
direction of the most massive node halo and the filament direction is almost
parallel, the two kinds of alignment signals are very similar. At very small
separation, however, the spin and direction of the most massive halo show
opposite alignment signal compare to that between the spin and filament. This
feature clearly indicates a {\it transition} of the 2-D collapse phase of the
filaments to the 3-D collapse phase of the cluster/node halos at small
separation.

\subsection{Shape-filament alignment}
\begin{figure}
  \plotone{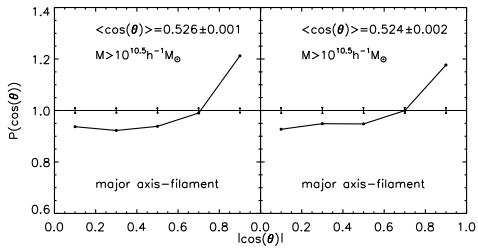}
 \caption{The probability distribution of the cosine of the angles between the
   halo  major axis  vectors and  the directions  of the  filaments  in Method
   \Rmnum{1} (left panel) and Method \Rmnum{2} (right panel).}
\label{major_filament}
\end{figure}
\begin{figure*}
\plotone{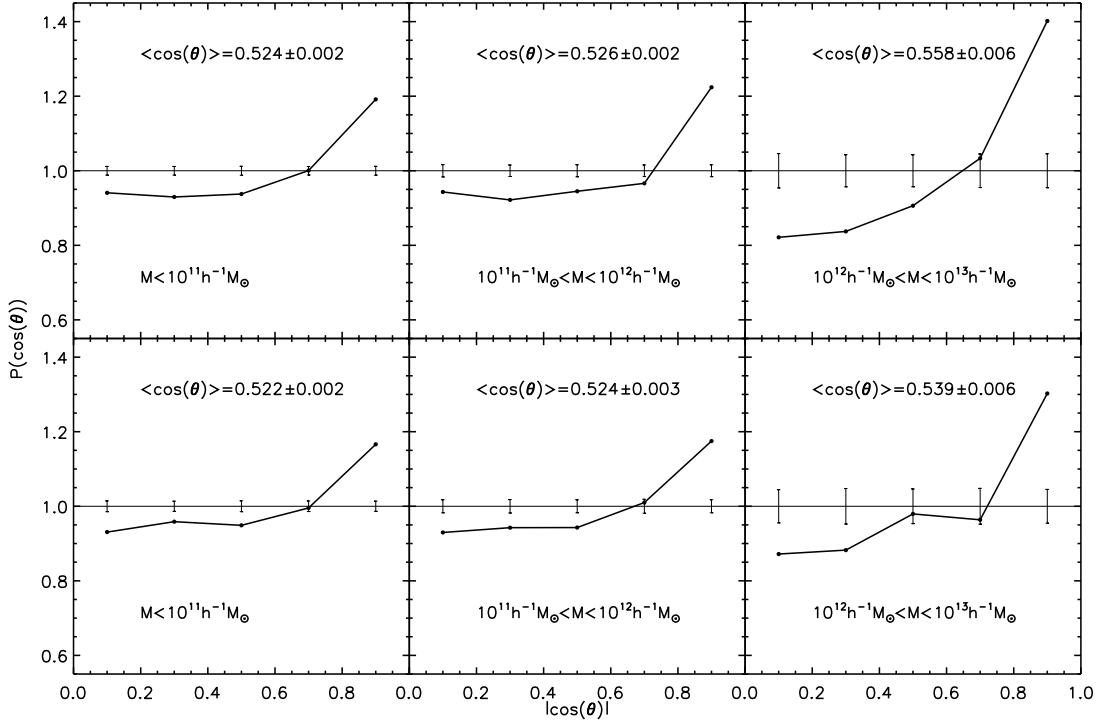}
\caption{Same as  Fig.~\ref{major_filament} but  for different mass  ranges of
  halos  in  the filaments  in  Method  \Rmnum{1}  (upper panels)  and  Method
  \Rmnum{2} (lower panels), as indicated.}
\label{major_massbin}
\end{figure*}
\begin{figure}
  \plotone{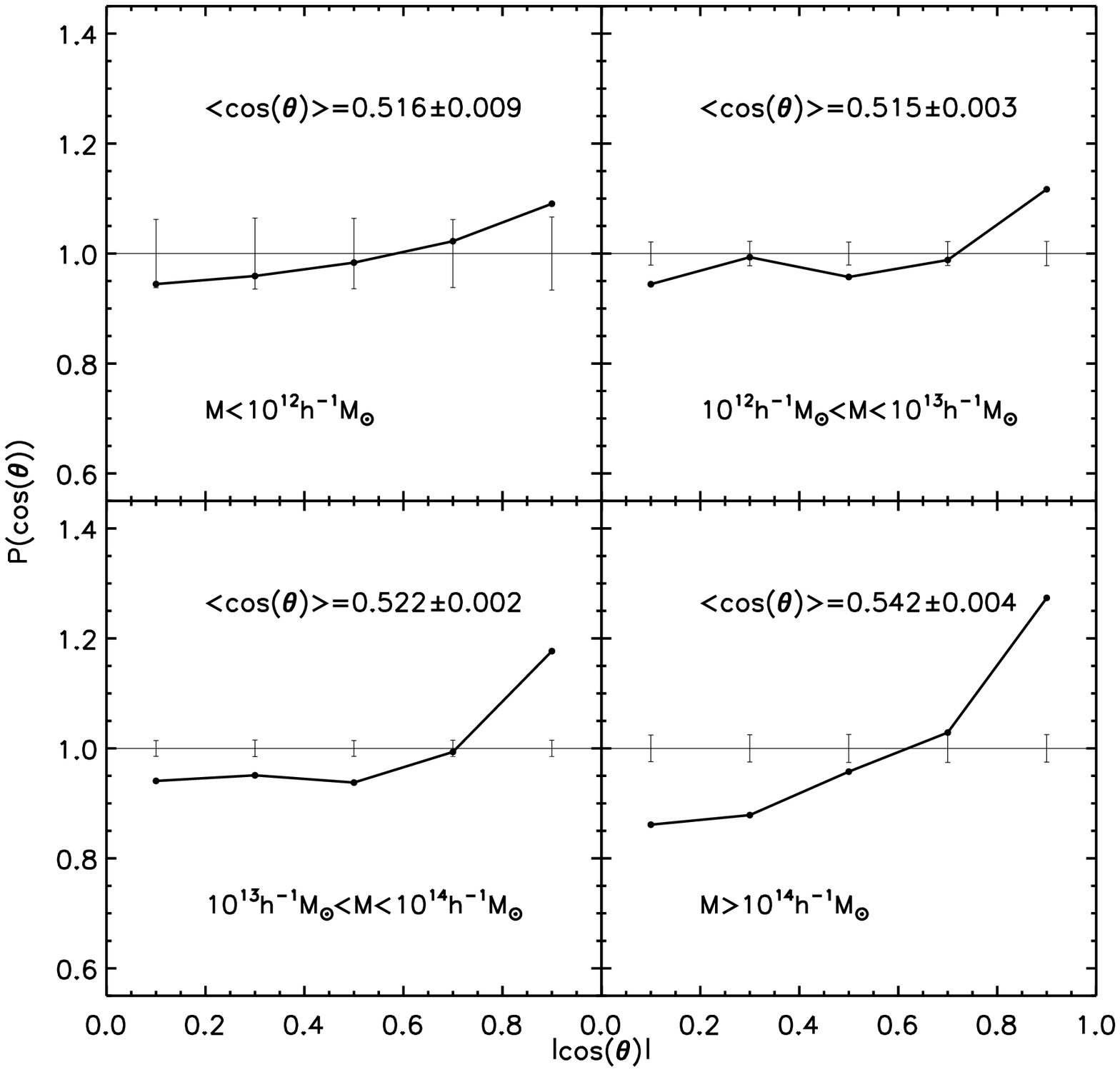}
\caption{Same as Fig.~\ref{major_filament} but for halos in different segments
  in Method  \Rmnum{2}, separated  according to the  mass of the  most massive
  halo in each segment. }
\label{major_masstop}
\end{figure}
\begin{figure}
  \plotone{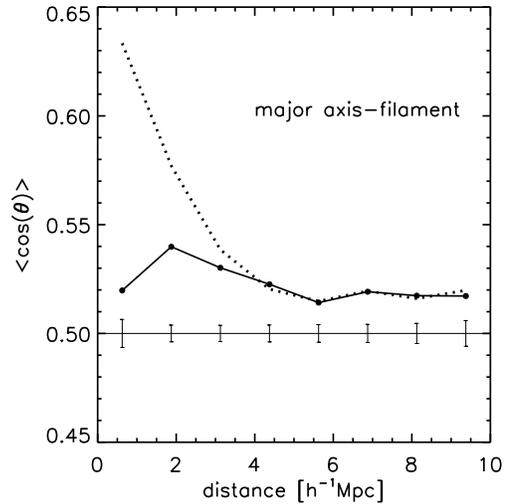}
  \caption{Mean cosine of the angle between shape and filament (solid line),
    between shape and direction of the most massive node halo (dotted line),
    as a function of the separations to the massive node halos in Method
    \Rmnum{2}.}
\label{major_distance}
\end{figure}
Next, we probe another important structural parameter, the orientation of the
halo shape with respect to the direction of the filament.  Similar to the last
section, we measure the alignment signals between the shapes and filaments for
filament halos.

In Fig.~\ref{major_filament}, we show the probability distribution of the
cosine of the angle between the halo major axis and the direction of the
filament. The left and right panels show the results for Methods \Rmnum{1} and
\Rmnum{2}, respectively. We find significant alignment signals with both
methods. In fact, the shapes of dark matter halos tend to be parallel to the
filaments.  Again, we use the average value of $\langle \cos(\theta) \rangle$
to quantify the strength of the alignment signal. From Method \Rmnum{1}, we
obtain an average cosine of $0.526\pm0.001$, whereas Method \Rmnum{2} results
in a slightly smaller value of $0.524\pm0.002$. Similar alignment trends are
reported in other recent studies (e.g., Altay et al. 2006; Arag\'on-Calvo et
al. 2007b; Hahn et al. 2007b).

In analogy to our investigation of spin-filament alignment, we now examine the
dependence of the alignment strength on mass and separation.
Fig.~\ref{major_massbin} shows the results for the filament halos in three
mass bins.  The upper and lower panels display results derived with Methods
\Rmnum{1} and \Rmnum{2}, respectively. The strength of the alignment grows
significantly with halo mass. Interestingly, the observed mass dependence
shows an opposite trend compared to the spin-filament alignment. These trends
agree well with results obtained by Hahn et al.~(2007b) and Arag\'on-Calvo et
al.~(2007b).

In Fig.~\ref{major_masstop}, we show the alignment signals for halos in
segments with most massive halos in four mass bins (note that this can only be
done for Method \Rmnum{2}). An obvious mass dependence of the alignment
signals is visible. Halos in segments with more massive node halos exhibit
stronger alignment signals.  The solid line in Fig.~\ref{major_distance}
displays the alignment signals between shape and filament for halos at
different separations to the most massive node halo. One can see a pronounced
distance dependence. Halos at smaller separations to the most massive node
halos, except at the smallest distance bin, tend to have stronger alignment
strength.  Similar to the spin of the halos, we also measure the alignment
signals between shape and direction of the most massive node halo. The results
are shown in Fig.~\ref{major_distance} as the dotted line. Obviously in this
measure the distance dependence is much enhanced and {\it monotonic}. Again,
this feature indicates the {\it transition} of the 2-D collapse phase of the
filaments to the 3-D collapse phase of the cluster/node halos at small
separation.  This distance dependence, if not restricted to the filament
members, is in general agreement with the alignment signals measured by
Faltenbacher et al.~(2008) for central and satellite halos as a function of
radius and Pereira et al. (2008) for substructures with their host halos.
\subsection{Spin-sheet and Shape-sheet alignment}
\begin{figure}
  \plotone{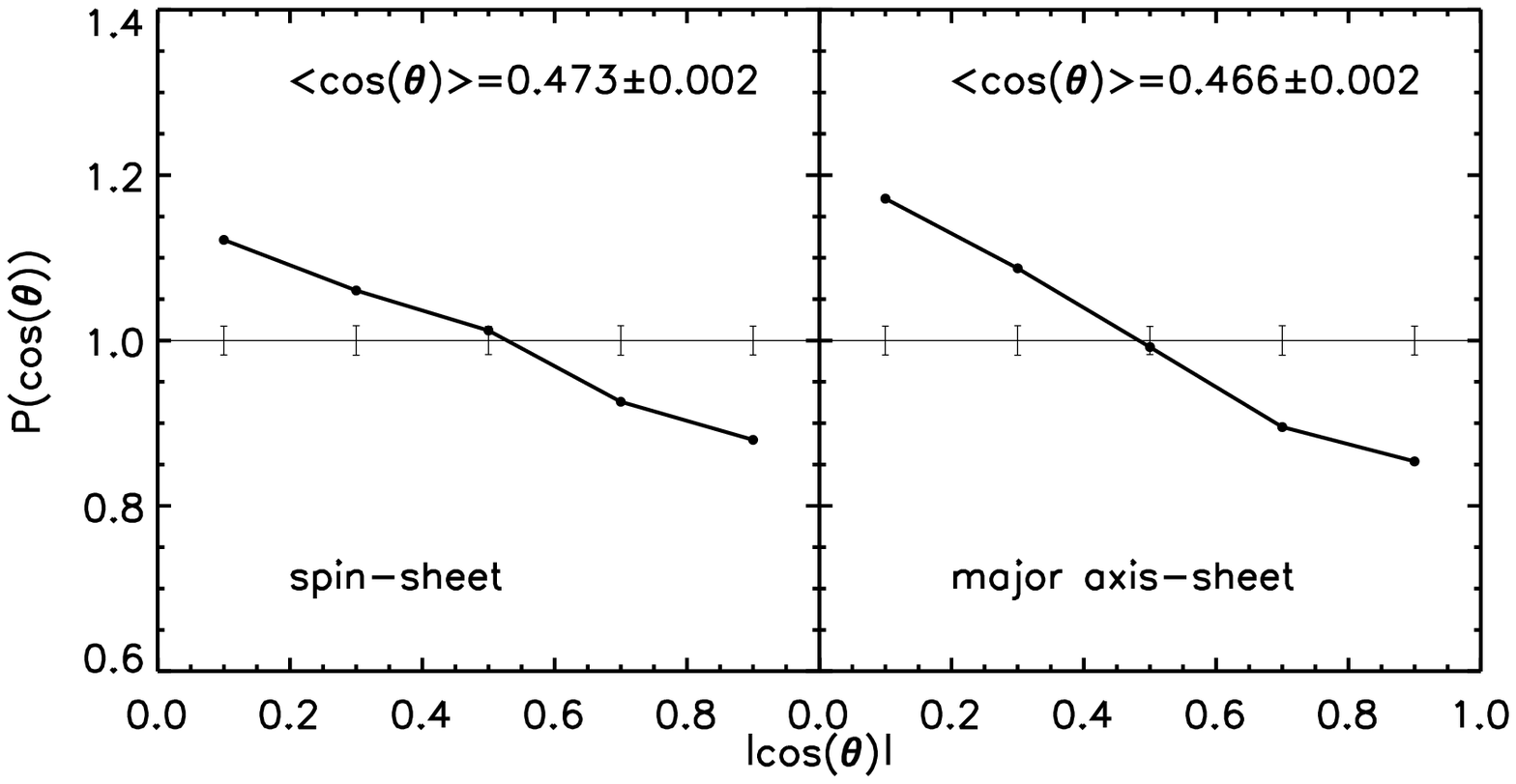}
\caption{Left panel: the  probability distribution of the cosine  of the angle
  between the halo angular momentum vector and the vector perpendicular to the
  sheet in  Method \Rmnum{1}. Right panel:  same as the left  panel but for
  the halo major axis vector.}
\label{sheet}
\end{figure}
\begin{figure*}
  \plotone{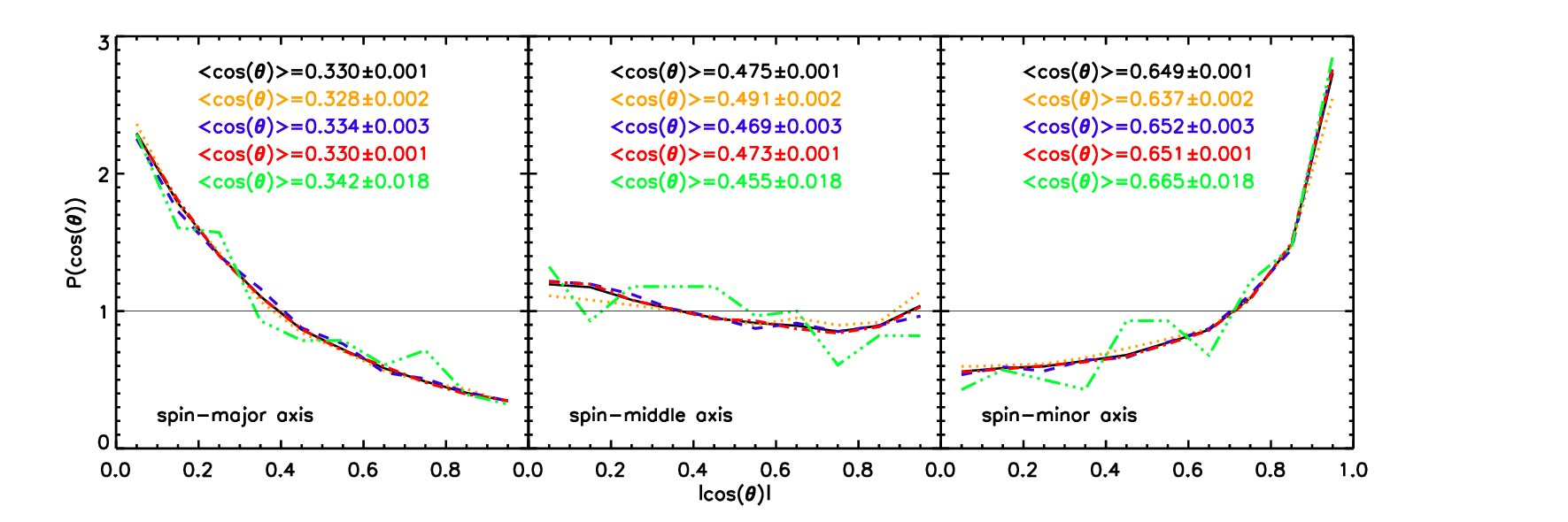}
  \caption{The probability distribution of the cosine of the angle between
    the angular momentum and major (left panel), middle (middle panel) and
    minor (right panel) axis vectors of the halos in clusters (orange dotted),
    filaments (red dash-dotted), sheets (blue dashed), and voids (green
    dash-dot-dot).  The black solid line indicates the values for all the
    halos in our halo catalog. }
\label{spin_major}
\end{figure*}
Method \Rmnum{1} differentiates between four cosmic environments: clusters,
filaments, sheets and voids. Filaments are distinguished by the condition that
their Hessian matrix has only one single positive eigenvalue, and the
corresponding eigenvector determines a unique direction. Sheets on the other
hand are defined by having only one single negative eigenvalue. The associated
eigenvector also determines a unique direction, which can be identified with
the normal to the sheet.
  
The alignment signal for the angle between the halo spin and the normal of the
sheet is shown in the left panel of Fig.~\ref{sheet}. We obtain an
anti-alignment signal, which means that there is a trend for sheet halos to
have their angular momentum vector parallel to the plane of the sheet. The
alignment strength, quantified by the average of the cosine is $\langle
\cos(\theta) \rangle=0.473\pm0.002$.  This alignment strength is in very good
agreement with that obtained by Arag{\'o}n-Calvo et al. (2007b).

The right panel of Fig.~\ref{sheet} shows the probability distribution of the
cosine of the angle between the halo major axis vector and the vector
perpendicular to the sheet. The major axes of halos in sheets are strongly
aligned with the sheet planes.  The average of the cosine is $0.466\pm0.002$.
A similar tendency has also been found in the shells of voids, as reported by
Brunino et al.  (2007).

\subsection{Spin-Shape alignment}
\begin{figure*}
  \plotone{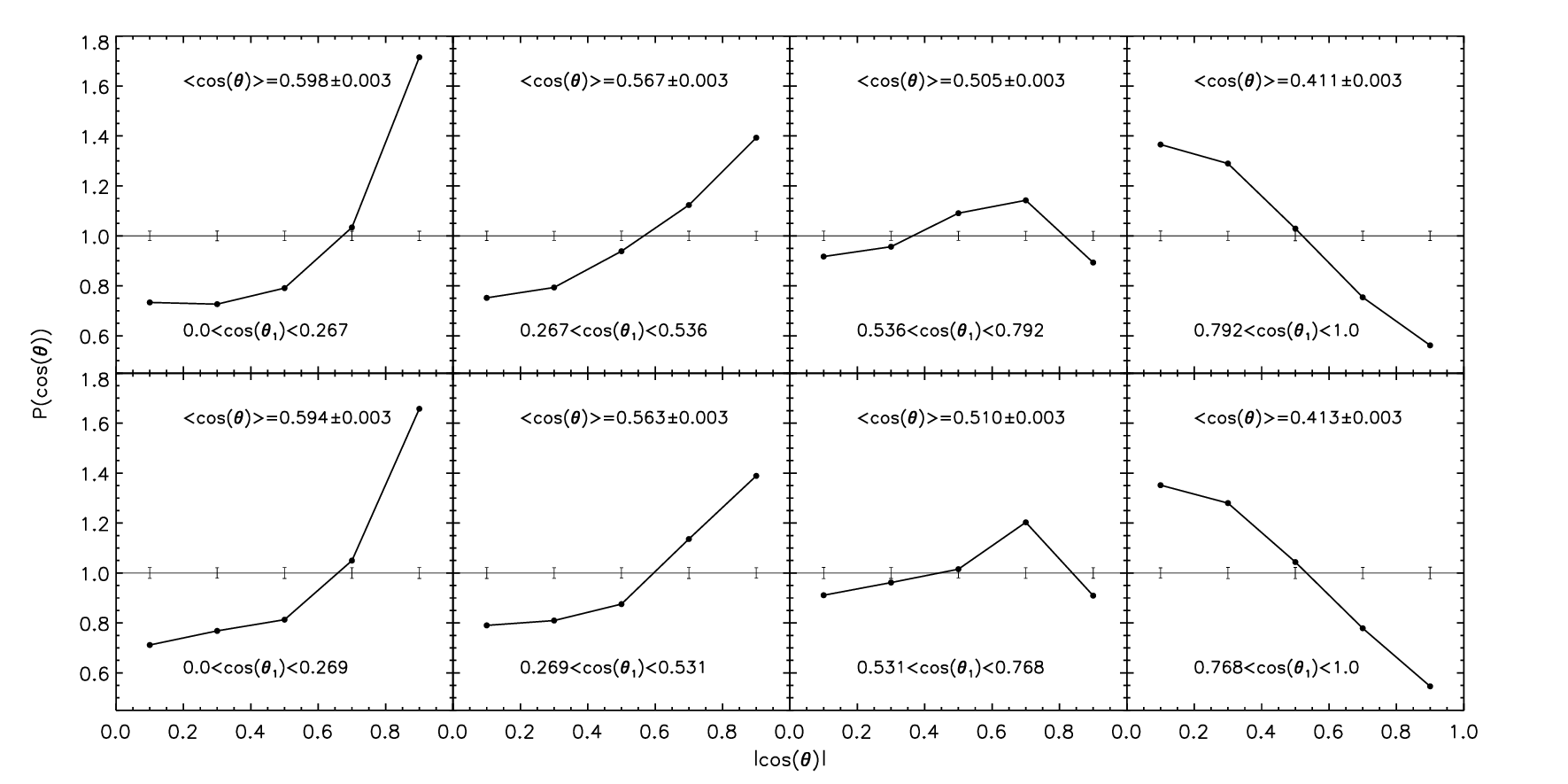}
  \caption{The probability distribution of the cosine of the angle between the
    halo spin axis vectors and the directions of the filaments in Method
    \Rmnum{1} (upper panels) and Method \Rmnum{2} (lower panels). Each panel
    from left to right corresponds to $1/4$ of the halos in the filaments,
    ranked according to the cosine of the angle between the halo major axis
    and the direction of the filaments, $\cos(\theta_1)$. In each panel, the
    corresponding range of $\cos(\theta_1)$ and the average value of $\langle
    \cos(\theta) \rangle$ are indicated. }
\label{spin-shape}
\end{figure*}
Having specified the spin and shape (major axis) alignments with respect to
the large-scale environments (filaments and sheets), we now proceed to examine
the spin-shape alignment within the halos themselves.  Fig.~\ref{spin_major}
shows the alignment signals between the spin-major axes (left panel),
spin-middle axes (middle panel) and spin-minor axes (right panel),
respectively. Measurements of these signals in different environments are
displayed with different line styles: results are given for clusters (orange
dotted), filaments (red dash-dotted), sheets (blue dashed), voids (green
dash-dot-dot) and for all halos (black solid). In each case, the average of
the cosine of the alignment angle is also indicated in each panel.

We find no significant environmental dependence of the alignment signal
between the spin and shape of halos.  The halo spin vector appears to be
preferentially perpendicular to the halo major axis, and has a strong tendency
to be parallel to the halo minor axis. This behavior is in good agreements
with previous findings obtained, e.g., by Faltenbacher et al. (2002), Bailin
\& Steinmetz (2005); Allgood et al.  (2006) and Bett et al. (2007).  We have
also separately investigated the alignment signal for halos with mass $\le
10^{11.0}\msun$ and $\ge 10^{12.5}\msun$, and basically found no dependence on
mass besides a marginal enhancement of the alignment signal for the $\ge
10^{12.5}$ halos.

According to our analysis above, both the spin and the major axes of halos are
preferentially aligned with the directions of the filaments or the planes of
the sheets.  On the other hand, within halos the spin axes are strongly
aligned with the {\em minor axes} of the halos. At first glance, these two
results seem contradictory.  In case of a perfect alignment between halo spin
and minor axes, an alignment between halo orientations and filaments would
cause an anti-alignment between spin axes and filaments.  

To have a better understanding of these two sets of `contradictory' results,
we perform the following additional test.  We first rank all the (member)
halos within the filaments according to the cosine of their angles between the
major axes and the filament directions, $\cos(\theta_1)$\footnote{To avoid the
  duplicated use of notation $\theta$, here we use $\theta_1$ to represent the
  angle between the major axis and the filament directions.}.  These halos are
then split into four bins with equal numbers according to the values of
$\cos(\theta_1)$.  We then measure the cosines of the angles between the spin
axis and the filament directions for the halos in each $\cos(\theta_1)$ bin.
The results are shown in Fig.~\ref{spin-shape}, with upper panels for Method
\Rmnum{1} and lower panels for Method \Rmnum{2}.  Each panel in
Fig.~\ref{spin-shape} from left to right corresponds to results for 1/4 of the
halos in the filaments within a different $\cos(\theta_1)$ range, as
indicated.

The most right hand panels of this figure clearly demonstrate that a {\em
  strong} alignment between the major axes and the directions of the filaments
produces an anti-alignment between the spin axes and the filaments. This
result is in agreement with simple geometric considerations. Note however,
theoretically a {\em strong} anti-alignment between the major axes and the
directions of the filaments, as addressed in the left hand panels of
Fig.~\ref{spin-shape}, {\it does not} guarantee an alignment between the spin
axes and the filaments. In the case of $\theta_1=90^o$ and perfect alignment
between minor axis and spin, the angle between the spin (minor) axis and the
filament direction can assume any value within $0-90^o$. The very strong
alignment signals apparent in the most left hand panels must be induced purely
by the influence of the filaments.  Combining the signals from the four
panels, one would obtain the alignment signals shown in
Fig.~\ref{spin_filament}. On the other hand, note the range of the
$\cos(\theta_1)$ in each panel, the average $\cos(\theta_1)$ for all halos in
the four panels is larger than $0.5$, again indicating an alignment between
the major axis and filament directions (as shown in
Fig.~\ref{major_filament}).  Thus we conclude that the two sets of alignments
are indeed not contradictory, and that the large-scale environment, i.e.~the
filaments and the sheets, can impact both the orientation and the spin of
halos while allowing still for an internal correlation of these quantities
within halos.

\subsection{Filament directions in Methods \Rmnum{1} and \Rmnum{2}}

\begin{figure}
  \plotone{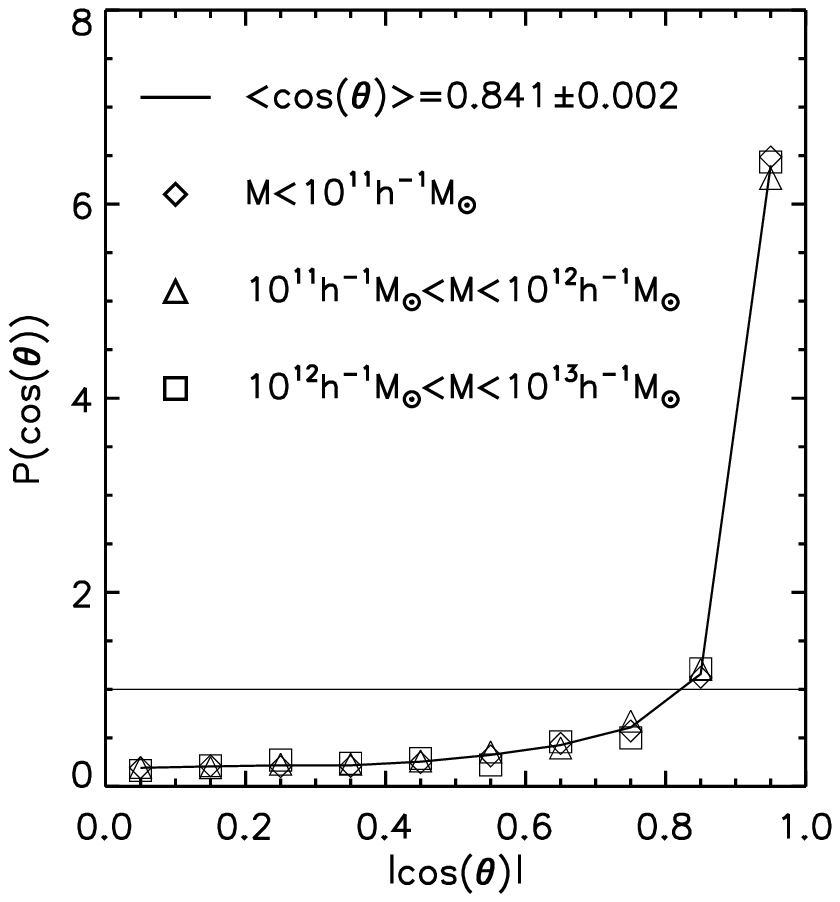}
\caption{The probability distribution of the cosine of the angle between the
directions of the filaments from Method \Rmnum{1} and the filament vectors
from Method \Rmnum{2} for all the filament halos (black solid) and different
mass range of halos. The average value of $\cos\theta$ of all the filament
halos is also shown.}
\label{fila1-fila2}
\end{figure}

Throughout the paper we find very good agreement between various alignment
signals for Methods \Rmnum{1} and \Rmnum{2}, however it is not necessory that
the same halos in the two methods have the same filament directions. In order
to make a more profound understanding of the agreement results of the two
methods, we compute the probability distribution $P(\cos\theta)$ of the angle
between the directions of the filaments from Method \Rmnum{1} and the filament
vectors from Method \Rmnum{2}. As shown in Fig.~\ref{fila1-fila2}, the
filament directions for the same halos in the two methods agree very well and
more than 60\% halos have $\theta\sim 0$. In addition, we did not find any
significant halo mass dependence for these filament angles. As we have also
checked the local density and distance to the most massive node halo of each
common halo, we do not find any significant dependence on this angle $\theta$.

\section{SUMMARY}\label{sec_summ}
Using the dark matter and halo distributions from a high resolution N-body
simulation, we have identified the filamentary structures in the cosmic web
using two different filament-finding algorithms. The first method is based on
the Hessian matrix of the density field (Method \Rmnum{1}), where the halos
are classified into four categories according to the signs of the three
eigenvalues of the Hessian matrix: clusters, filaments, sheets and voids.  The
advantage of this method is that the large-scale environments of the halos can
be characterized unambiguously, and that there is only one free parameter in
the analysis.  However, this method needs detailed information about the local
density field, which is difficult but not impossible to obtain based on
observations. Indeed, some recent studies are devoted to density field
reconstructions based on galaxy and halo distributions (e.g. Erdogdu et
al. 2006 from 2MASS Redshift Survey; and Wang et al.  2009a from halo
distribution in simulations).  

On the other hand, Method \Rmnum{2} directly uses the distribution of halos.
Thus it eliminates the need for a reconstruction of the local mass density
field, at the price of a larger number of tunable parameters.  Filamentary
structures are here traced by connecting single segments which are identified
according to their relative mass over-density. In this study we were not
interested in explicitly reconstructing long coherent filaments, rather we
only explore the orientations of the segments relative to the halos
orientations associated with them. The main shortcomings of Method \Rmnum{2}
are the relatively large number of parameters and a lack of sensitivity for
the detection of less prominent filaments. Nevertheless, Method \Rmnum{2} is
easy to implement on top of observational data, including also galaxy group
catalogs such as that of Yang et al.~(2007). This makes it a highly useful
approach in practice, provided its results are consistent with those obtained
with Method \Rmnum{1}. This is indeed the case, as we have demonstrated in
this study. Moreover in Method \Rmnum{2}, halos are automatically grouped into
filaments, and we know what halo belongs to what filament.

Based on the classification of the large-scale environment around halos that
we obtained, we examined the spin-filament and the shape-filament alignments,
as well as the corresponding quantities for halos and sheets.  Overall, we
find a tendency for halo spins and shapes to be oriented parallel to filaments
and perpendicular to the normal vector of sheet like structures.  This trend
becomes obvious and is clearly significant when compared to results from
isotropic (randomized) orientations.  Both the spin and shape alignment
strengths are stronger for halos close to more massive node (cluster) halos,
and at smaller separations to the node halos.  Yet, the spin and shape
alignment strengths show an opposite mass dependence: smaller halos display
stronger (weaker) spin (shape) alignments with the filaments.

For the halos in different environments, we have also investigated their
intrinsic spin-shape alignment. We found that the spin axes are preferentially
aligned with the minor axes of halos, and are perpendicular to the major axes,
independent of the halo environments. At first glance, this result seems to be
at odds with our finding that {\em both} the spin and major axes tend to align
with the filament directions. However, there is not really a contradiction
here, because the alignments {\em are not perfect} in both cases. In fact, the
distributions of the alignment angles are sufficiently broad that both of
these seemingly contradictory alignment signals can coexist.

Our results are in good agreement with recent N-body studies where different
filament finding methods have been employed.  Hahn et al.~(2007b) applied a
Hessian matrix approach to the gravitational potential field (instead of the
density field), and also found an opposite mass dependence for the alignment
strengths of spin- and shape-filament alignments.  Moreover, they found that
the spin axes of halos with mass $\ge 10^{13}\msun$ are preferentially
perpendicular to the filament directions, which we cannot confirm in this
study because of the limited volume of our simulation. A similar behavior has
been detected by Arag\'on-Calvo et al. (2007b) using a multi-scale morphology
filter method for the classification of the large-scale environments.

Quite interestingly, we find in Method \Rmnum{2} according to the spin and
shape alignment signals of the halos in the filaments, there is a transit
influence scale of large scale environments: from the 2-D collapse phase of
the filament to the 3-D collapse phase of the cluster/node halo at small
separation to the most massive node halo.

The general trends found from our N-body simulation reveal a substantial
interplay between the large-scale environments and the internal properties of
the dark matter halos.  It would be interesting to see whether similar trends
can be observed in real galaxy samples (e.g., Lee \& Pen 2001; Lee \& Erdogdu
2007). The two methods outlined in this study should be straightforwardly
applicable to observational data sets (e.g.~SDSS).  This should provide for
interesting tests of galaxy formation models, and of cold dark matter
cosmologies.

\acknowledgements

We are grateful to the anonymous referee for useful and insightful comments
that greatly helped to improve the presentation of this paper.  The simulation
was done at Shanghai Supercomputer Center. This work was supported by 973
Program (Nos. 2007CB815401, 2007CB815402), 863 program (2006AA01A125), the CAS
Knowledge Innovation Program (Grant No. KJCX2-YW-T05) and grants from NSFC
(Nos.  10533030, 10633049, 10821302, 10873027, 10925314).

\end{document}